\documentclass[sigconf]{acmart}

\usepackage[T1]{fontenc}
\usepackage{multirow}
\usepackage{threeparttable}
\usepackage[nointegrals]{wasysym} 
\usepackage{tikz}
\usepackage{xurl}
\usepackage{tabularx}
\usepackage{booktabs,xcolor,colortbl}
\AtBeginDocument{%
  \everydisplay\expandafter{\the\everydisplay
    \abovedisplayskip=4pt plus 1pt minus 1pt
    \belowdisplayskip=4pt plus 1pt minus 1pt
    \abovedisplayshortskip=2pt plus 0.5pt minus 1pt
    \belowdisplayshortskip=4pt plus 1pt minus 1pt
  }%
}

\definecolor{operationgroup}{RGB}{226,241,250}
\newenvironment{packeditemize}{
	\begin{list}{$\bullet$}{
	\setlength{\labelwidth}{8pt}
	\setlength{\itemsep}{0pt}
	\setlength{\leftmargin}{\labelwidth}
	\addtolength{\leftmargin}{\labelsep}
	\setlength{\parindent}{0pt}
	\setlength{\listparindent}{\parindent}
	\setlength{\parsep}{0pt}
	\setlength{\topsep}{3pt}}}{\end{list}}

\usetikzlibrary{arrows.meta,positioning,fit,calc}

\newcommand{\fullcircle}{\textcolor{teal!58}{\CIRCLE}}
\newcommand{\halfcircle}{\textcolor{orange}{\LEFTcircle}}
\newcommand{\emptycircle}{\textcolor{violet!78}{\Circle}}

\AtBeginDocument{%
  }

\setcopyright{acmlicensed}
\copyrightyear{2018}
\acmYear{2018}
\acmDOI{XXXXXXX.XXXXXXX}
\acmConference[Conference acronym 'XX]{Make sure to enter the correct
  conference title from your rights confirmation email}{June 03--05,
  2018}{Woodstock, NY}
\acmISBN{978-1-4503-XXXX-X/2018/06}

\begin{document}

\title[AMLBench: A Real-Incident Benchmark]{AMLBench: A Real-Incident Benchmark for Agentic Blockchain Anti-Money Laundering}



\author{Qishuang Fu}
\affiliation{%
  \institution{Monash University}
  \city{Melbourne}
  \country{Australia}}

  \author{Hang Zheng}
\affiliation{%
  \institution{Monash University}
  \city{Melbourne}
  \country{Australia}}

    \author{Yuchen Lei}
\affiliation{%
  \institution{Independent researcher}
  \city{ }
  \country{ }}

  \author{Yuexin Xiang}
\affiliation{%
  \institution{Monash University}
  \city{Melbourne}
  \country{Australia}}

    \author{Qin Wang}
\affiliation{%
  \institution{CSIRO}
  \city{Sydney}
  \country{Australia}}

    \author{Yuanzhe Zhang}
\affiliation{%
  \institution{Nanyang Technological University}
  \country{Singapore}}
  
    \author{Shiping Chen}
\affiliation{%
  \institution{CSIRO}
  \city{Sydney}
  \country{Australia}}

  \author{Andreas Deppeler}
\affiliation{%
  \institution{Monash University}
  \city{Selangor}
  \country{Malaysia}}

  \author{Joseph K. Liu}
\affiliation{%
  \institution{Monash University}
  \city{Melbourne}
  \country{Australia}}

    \author{Ziyao Liu}
\affiliation{%
  \institution{Nanyang Technological University}
  \country{Singapore}}

    \author{Tsz Hon Yuen}
\affiliation{%
  \institution{Monash University}
  \city{Melbourne}
  \country{Australia}}

\renewcommand{\shortauthors}{Anonymous Authors}

\begin{abstract}

Tracing illicit funds on public blockchains remains largely \textit{manual}.
As funds move across accounts and undergo asset conversions, they become mixed with other inflows, requiring repeated queries and expert judgment to trace.
Large language model (LLM) agents 
offer a way to
automate this process, but their ability to reconstruct case-related transfer networks from limited clues remains unclear.
To bridge this gap, we present \textit{AMLBench}, an 
anti-money laundering (AML) benchmark built from 24 real-world cases and 5.05 million Ethereum transfer records.
For each case, agents start from limited clues and explore these transfer records through composable transaction-query and fund-flow analysis tools, deciding which accounts and branches to pursue.
The final tracing results are evaluated against case annotations, while tool interaction logs reveal how agents discover and select relevant transfers and at what resource cost.
We also develop \textit{AMLAgent}, which integrates fund-tracing guidance and query feedback to dynamically guide investigative decisions.
With GPT-6 Astra, AMLAgent achieves macro averages of 80.82\% transfer-level precision and 84.73\% query coverage across 24 real-world cases.
Further analysis of the large-scale Bybit case reveals substantial differences in transfer retention and token usage across models, even with similar submission precision.
AMLBench establishes a reproducible benchmark for evaluating both tracing outcomes and investigative processes, laying the groundwork for more reliable and efficient autonomous AML investigations.

\end{abstract}

% ...

\ccsdesc[500]{Applied computing~Computer forensics}
\ccsdesc[300]{Computing methodologies~Intelligent agents}

\keywords{Blockchain Forensics, AML, LLM Agents, Benchmark}


\maketitle

\section{Introduction}
\label{sec:intro}

\begin{figure}[!]
    \centering
    \includegraphics[width=0.80\linewidth]{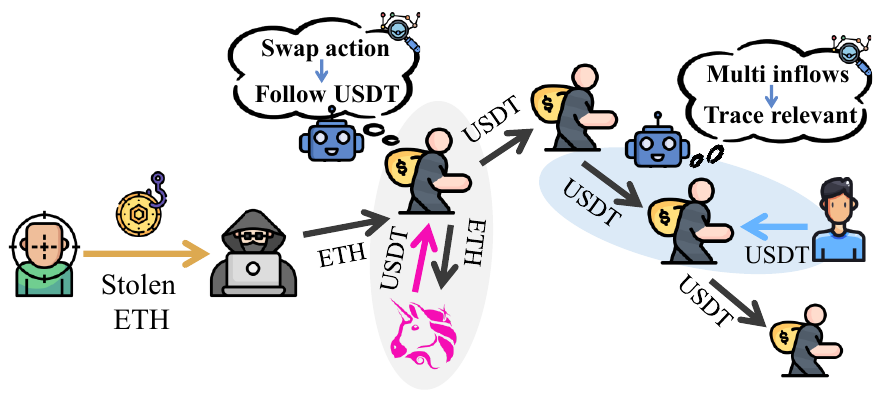}
    \vspace{-0.1in}
    \caption{Agentic AML investigations in Web3 finance.}
    \vspace{-0.15in}
    \label{fig:motivation}
\end{figure}

Web3 finance relies on blockchains such as Ethereum, where publicly accessible transaction records provide a basis for tracing illicit funds~\cite{zhou2023sok}.
Consider a cryptocurrency theft: the thief moves stolen funds through multiple accounts and asset swaps to obscure their origin~\cite{fatf2020virtual}.
As illustrated in Fig.~\ref{fig:motivation}, investigators follow stolen funds through asset swaps and successive accounts, where mixed inflows and branching paths complicate the identification of relevant transfers.
Reconstructing these fund movements and identifying their eventual destinations from on-chain records is a central task in anti-money laundering (AML) investigations.
Such investigations play a critical role in exposing illicit financial flows and safeguarding the integrity of Web3 finance~\cite{bellei2024shape}.

Prior work has explored automating blockchain investigations by identifying suspicious activity using transaction and graph features~\cite{weber2019anti,farrugia2020detection,bellei2024shape} and tracing funds from known leads through heuristic rules and graph search~\cite{yousaf2019tracing,wu2023tracer}.
However, these approaches rely on predefined detection criteria or tracing rules, making it difficult to adapt investigative decisions as new leads emerge.
More recently, large language model (LLM) agents have enabled iterative investigations by querying blockchain data and using the results to decide what to investigate next~\cite{yao2023react}.
For example, RiskTagger~\cite{lin2026risktagger} uses an LLM-guided tracing process to expand fund-flow graphs based on on-chain evidence.
Yet existing work primarily focuses on developing tracing methods, with limited benchmark support for systematically evaluating the accuracy and efficiency of autonomous fund tracing across real-world cases.





A key challenge in evaluating autonomous fund tracing is that an agent's decisions about which addresses to follow and how far to trace them can leave entire branches of the transfer network unexamined.
This raises a central question: \textbf{How accurately and completely can an agent reconstruct an incident's transfer network from a few initial clues via autonomous queries?}


To answer this question, we need a \textit{benchmark for real AML investigations} that evaluates agents'
ability to reconstruct case-related fund flows under standardized conditions while recording their investigation processes.
However, constructing such a benchmark presents two challenges.

\begin{packeditemize}
    \item \textbf{C1: How can real cases be transformed into investigation tasks that support autonomous exploration?}
    Fund tracing is an iterative process in which agents discover new addresses and transfers from available evidence and decide what to investigate next~\cite{yu2026locard}.
    This calls for tasks that support iterative querying while preserving other transactions and fund-flow branches for exploration.
\item \textbf{C2: How can evaluation remain fair and comparable while allowing different investigation strategies?}
    Starting from the same initial clues, agents may follow different investigation paths or stop at different points, leading to different outcomes.
    A fair evaluation therefore accommodates different valid paths to the same goal~\cite{zhou2024webarena} while measuring tracing accuracy and coverage alongside query costs under consistent conditions~\cite{kapoor2025ai}.
\end{packeditemize}

To address these challenges, we introduce \textbf{AMLBench}, a benchmark for evaluating autonomous fund tracing in real-world AML cases.
For \textbf{C1}, we formulate each case as an investigation task that preserves surrounding transaction activity and provides composable atomic query operations for agents to explore fund flows from limited initial clues.
For \textbf{C2}, an automated evaluation pipeline compares agents' tracing results against case annotations and records tool interactions and resource costs, enabling consistent assessment of both investigation outcomes and processes.
To evaluate different LLM backbones on AMLBench, we also develop \textbf{AMLAgent}, which progressively traces case-related funds through iterative tool calls.


Our work makes the following contributions:
\begin{packeditemize}  
    \item \textbf{AMLBench} (\S\ref{sec-amlbench}).
    We establish a reproducible benchmark based on 24 real-world cases, enabling autonomous fund tracing under consistent investigation conditions with transfer-level evaluation of outcomes and processes.
    \item \textbf{AMLAgent} (\S\ref{sec-amlagent}).
    We develop a reusable fund-tracing agent that integrates tracing guidance and query feedback to support autonomous investigations, providing a baseline for comparing different LLM backbones.

    \item \textbf{Evaluation}  (\S\ref{sec-evalua}).
    We systematically evaluate autonomous fund tracing across real-world cases and different LLM backbones, revealing the capabilities and failure modes of current agents. This establishes a foundation for understanding the reliability of LLM-driven AML investigations and identifying critical challenges to practical deployment. 
\end{packeditemize}

\section{Related Work}
\label{sec-rw}

We briefly review the most closely related lines of work here and provide detailed comparisons in Appendix~\ref{app:related_work}.

\subsection{Blockchain Anti-Money Laundering}

Existing blockchain AML research primarily focuses on suspicious activity detection and fund tracing.
One line of work uses transaction features and graph structure to identify suspicious transactions~\cite{weber2019anti}, accounts~\cite{farrugia2020detection}, or laundering subgraphs~\cite{xiang2022leveraging,bellei2024shape}.
These methods help identify potential investigation targets, but typically formulate the problem as detection over given data, without fully capturing how real investigations proceed from limited clues through interactive queries and the discovery of new targets.
Another line of work traces funds from initial clues using taint propagation~\cite{wu2023towards}, relevance-guided graph search~\cite{wu2023tracer}, or transaction-linking rules~\cite{fu2026flowshield,fu2026mixguard}.
These methods construct fund-flow subgraphs, but predefined tracing criteria can cause them to miss relevant branches or explore irrelevant ones.
This motivates interactive tracing tasks that allow agents to adapt their investigations as new evidence emerges.

\subsection{Agentic Blockchain Forensics}

Recent research applies LLMs and agents to blockchain transaction analysis and autonomous investigation~\cite{yu2026locard}.
Lei et al.~\cite{lei2025large} explore LLMs' ability to understand and interpret transaction graphs.
RiskTagger~\cite{lin2026risktagger} extracts clues from public incident reports and iteratively expands fund-flow graphs using on-chain evidence to support forensic investigations.
These studies demonstrate the potential of LLM agents to advance blockchain investigations through evidence gathering and sequential decision-making.
However, their task settings and evaluation criteria differ, making investigative performance difficult to compare.
We therefore introduce AMLBench to evaluate different agents' fund-tracing capabilities on real incidents under consistent conditions.

\begin{figure*}[t]
    \centering
    \includegraphics[width=0.94\textwidth]{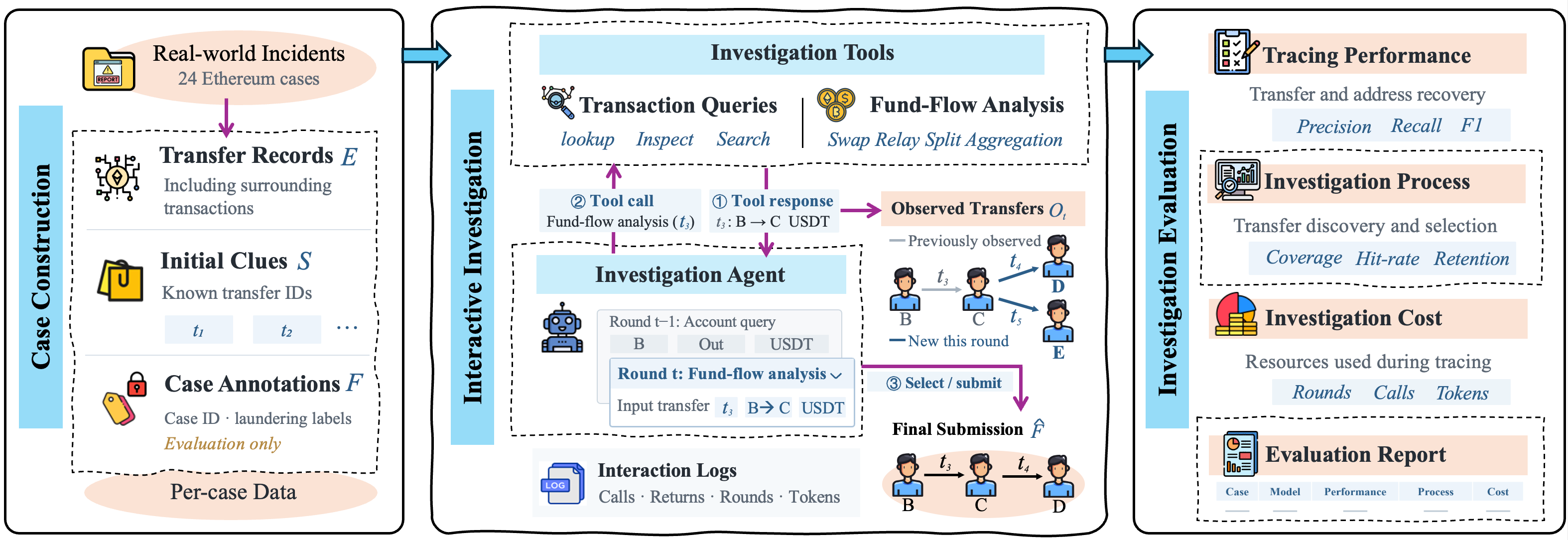}
    \vspace{-0.09in}
    \caption{AMLBench Overview.}
    \label{fig:amlbench_overview}
        \vspace{-0.1in}
\end{figure*}

\section{Preliminaries}

We introduce the basic concepts for incident-level fund tracing.

\subsection{Blockchain Transactions}

\textbf{Token transfer.}
On smart contract-enabled blockchains such as Ethereum~\cite{buterin2014ethereum}, a single transaction can generate multiple token transfer records.
Each transfer has a unique identifier and attributes $(\tau,u,v,a,q)$, denoting its transaction hash, sender, recipient, asset type, and amount, respectively.
The set of transfers associated with transaction $\tau$ is denoted by $\operatorname{Tx}(\tau)$.
Transfers within the same transaction may involve different senders, recipients, or assets, while sharing the transaction hash and timestamp.

\smallskip
\noindent\textbf{Fund flow.}
Given a transfer set $E$, its fund-flow graph is defined as a directed temporal multigraph $G=(V,E)$, where nodes represent addresses and edges represent asset transfers, retaining their assets, amounts, timestamps, and transaction hashes~\cite{wu2023tracer}.
For a target transfer from $u$ to $v$, earlier transfers into $u$ are candidate upstream transfers, while later transfers out of $v$ are candidate downstream transfers~\cite{fu2026flowshield}.
Expanding upstream traces potential funding sources, while expanding downstream follows potential fund destinations.

\subsection{Incident-Level Fund Tracing}



Incident-level fund tracing is a core task in anti-money laundering investigations.
Starting from known clues about a specific incident, it traces the sources, paths, and destinations of related funds to help analyze laundering activities and relationships among involved accounts~\cite{yu2026locard,lin2025risktagger}.
We model this task as an interactive process that progressively reconstructs the incident's fund-flow network through iterative queries.

\smallskip
\noindent\textbf{Task Setting.}
For an incident, let $G=(V,E)$ be the fund-flow graph formed by the transfer records available for querying, where $E$ includes both incident-related and background transfers.
The initial clues consist of a small set of known transfers $S\subseteq E$, and the set of transfers labeled as incident-related is denoted by $F\subseteq E$.
Starting from $S$, the agent progressively retrieves additional records through queries.

\smallskip
\noindent\textbf{Interactive Tracing.}
Let $O_t\subseteq E$ denote the set of transfers retrieved up to step $t$.
Based on $S$ and $O_t$, the agent decides which addresses to query next, whether to trace upstream or downstream, and when to stop.
At the end of the investigation, the agent submits a set of transfers detected as incident-related, denoted by $\widehat{F}\subseteq E$, with the goal of identifying $F$ as accurately as possible.





\section{AMLBench}
\label{sec-amlbench}


We propose AMLBench (Figure~\ref{fig:amlbench_overview}), a benchmark for evaluating autonomous blockchain fund tracing. 
We describe its three core components: task construction from real cases, an interactive investigation environment with composable tools, and evaluation of both tracing outcomes and investigation processes.

\subsection{Case Construction}

\noindent\textbf{Case Sources.}
We build AMLBench from 24 cryptocurrency security incidents between 2018 and 2025.
The data come from two studies of cryptocurrency laundering~\cite{wu2023towards,fu2026flowshield}, which used public reports and known incident addresses to trace funds and collect on-chain records.
These datasets provide the transfer records for constructing investigation tasks and the case annotations for evaluating agents' tracing performance.

\vspace{2pt}
\noindent\textbf{Data Composition.}
Each case consists of on-chain transfer records, case annotations, and initial investigation clues.
\ding{182}~Each record in the transfer set $E$ includes the sender, recipient, asset type, amount, timestamp, and transaction hash.
We retain surrounding transactions involving incident accounts from the source data, so agents need to decide which transfers to follow based on the fund flows they observe.
\ding{183}~Case annotations specify which case each transfer belongs to and whether it is labeled as laundering-related.
We use these annotations to check agents' submitted tracing results.
The laundering-related labels used for evaluation are not shown to agents.
\ding{184}~The initial clues $S\subseteq E$ consist of a small set of known transfers provided to the agent at the start of an investigation.
They include transfers into incident accounts and a downstream transfer when available.
This setup simulates real investigations in which investigators start from known transactions and gradually trace the sources, paths, and destinations of funds.


\smallskip
\noindent\textbf{Dataset Statistics.}
Table~\ref{tab:dataset_summary} summarizes the overall scale of AMLBench and the variation across cases.
The 24 cases contain approximately 5.05 million transfer records, ranging from 32 to 2.62 million records per case.
The table also reports per-case statistics on addresses, laundering-related transfers, data time spans, and initial clues to describe the investigation scope and starting conditions of different tasks.
Full statistics with 24 cases are in Appendix Table~\ref{tab:case_statistics}.

\smallskip
\noindent\textbf{Case Diversity.} The proportion of labeled relevant transfers varies substantially across cases, from 0.7\% in Bitpoint and 7.6\% in Bybit to 89.0\% in LCX and 90.0\% in Uranium Finance. Cases with sparse relevant transfers require agents to distinguish target fund flows from extensive background activity, while denser cases present different tracing challenges. This diversity enables AMLBench to evaluate investigations under varying evidence densities.

\begin{table}[H]
    \centering
 \small
    \renewcommand{\arraystretch}{1}
    \begin{threeparttable}
    \caption{AMLBench summary.}
    \label{tab:dataset_summary}
         \vspace{-0.1in}
    \begin{tabular*}{0.85\columnwidth}{@{\extracolsep{\fill}}cccc@{}}
        \toprule
        \textbf{Per-case statistic}  & \textbf{Min}  & \textbf{Max}    & \textbf{Mean} \\
        \midrule
        Transfers           & 32   & 2.62M  & 210.6K \\
        Addresses           & 15   & 578.0K & 28.9K \\
        Laundering transfers & 20  & 431.7K & 28.8K \\
        Initial clues       & 1    & 93     & 14.0 \\
        \bottomrule
    \end{tabular*}
    \begin{tablenotes}[flushleft]
        \footnotesize
        \item 24 cases spanning 2018--2025, with 5.05M transfers in total.
    \end{tablenotes}
    \end{threeparttable}
\end{table}

\subsection{Interactive Investigation}
Building on the case data, AMLBench organizes fund tracing as a multi-turn interaction between an agent and the investigation environment.
Starting from the initial clues, the agent calls tools to query transaction records and uses the returned information to decide which accounts to examine next, whether to trace funds back to their sources or follow subsequent transfers, and when to end the investigation.
This setup simulates how investigators query records, assess the results, and adjust their tracing plans, allowing us to evaluate agents' ability to gather information and advance fund-tracing investigations autonomously.


\vspace{2pt}
\noindent\textbf{Task Initialization.}
To initialize an investigation, the benchmark runner loads the transfer set $E$ for a single case.
This set includes incident-related transfers and surrounding transactions.
The agent receives the case identifier, blockchain, investigation time window, and identifiers of the transfers in the initial clue set $S$.
It also receives descriptions of the available query operations and their parameters, and uses these operations to inspect transfer details and account inflows and outflows as needed to trace fund flows.

\vspace{2pt}
\noindent\textbf{Query--Response Interaction.}
To support agents continue tracing funds as they discover new transfers and accounts, AMLBench provides a set of composable investigation operations covering transaction record queries and fund-flow analysis.
As shown in Table~\ref{tab:investigation_operations}, record queries retrieve account activity and transfers within transactions, while fund-flow analysis examines asset changes and relationships between inflows and outflows to suggest candidate swaps, splits, and aggregations.
Returned addresses, transaction hashes, and transfer identifiers can be reused in subsequent queries to retrieve further records and verify candidate fund-flow links.

In each decision round, the agent selects one or more operations from the available set $\mathcal T$ and specifies their parameters based on the initial clues and previous interactions.
A call is represented as $c=(f,\mathbf p)$, where $f\in\mathcal T$ is the selected operation and $\mathbf p$ contains its parameters, such as transfer identifiers, addresses, and time ranges.
The environment executes $c$ on the case data $E$ to produce a response $r=\operatorname{Execute}(E,c)$ containing transfer records or candidate fund-flow links, with pagination and status information as needed.


\begin{table}[!]
    \centering
    \small
    \renewcommand{\arraystretch}{1.05}
    \caption{Operations in AMLBench.}
    \label{tab:investigation_operations}
    \vspace{-0.1in}

    \begin{threeparttable}
    \begin{tabular*}{0.95\columnwidth}
    {@{\extracolsep{\fill}}>{\columncolor{operationgroup}}ccll@{}}
        \toprule
        \multicolumn{1}{c}{} &
        \multicolumn{1}{c}{\textbf{Operation}} &
        \multicolumn{1}{c}{\textbf{Input}} &
        \multicolumn{1}{c}{\textbf{Output}} \\
        \cmidrule{2-4}
        & Transfer lookup  & Transfer ID         & Transfer details \\
        & Account activity & Address, direction  & In/out transfers \\
        & Tx inspection    & Tx hash             & Tx transfers \\
        \multirow{-4}{*}{\rotatebox{90}{\scriptsize\bfseries Tx qry.}}
        & Transfer search  & Address/asset, time & Matched transfers \\
        \cmidrule{2-4}
        & Swap        & Same-tx assets    & Asset change \\
        & Relay       & Inflow + outflow  & Single outflow \\
        & Splitting   & Inflow + outflows & Multiple outflows \\
        \multirow{-4}{*}{\rotatebox{90}{\scriptsize\bfseries FF anal.}}
        & Aggregation & Inflows & Aggregation links \\
        \bottomrule
    \end{tabular*}
    \begin{tablenotes}[flushleft]
        \footnotesize
    \item[] \textbf{Tx qry.} denotes transaction query;
\textbf{FF anal.} denotes fund-flow analysis.
    \end{tablenotes}
    \end{threeparttable}
      
\end{table}

Query responses guide subsequent operations and parameters, allowing the agent to adjust its investigation path.
For example, a possible swap found through an account query may prompt fund-flow analysis and queries of related transfers for check.
If the records support a swap from ETH to USDT, the agent can follow USDT outflows, and other incoming transfers may prompt upstream tracing.


To relate the query process to the final submission, let $R_t\subseteq E$ denote the transfers returned as records or referenced by candidate identifiers across all calls in round $t$.
The cumulative observed set is updated as $O_t=O_{t-1}\cup R_t$.
The case data $E$ remain fixed, and $O_t$ includes transfers encountered in responses, whether or not the agent considers them relevant to the case.
At termination after $T$ decision rounds, we retain $O_T$ alongside any final submission $\widehat F$. 

\vspace{2pt}
\noindent\textbf{Submission and Recording.}
When the agent decides to stop, it submits $\widehat F$, the set of transfers it considers relevant, using unique identifiers and giving a reason for stopping.
AMLBench records whether the run ends with a submission, reaches the round limit without submitting, or fails during execution.
Tool calls, parameters, responses, and resource usage are logged throughout the investigation.
These logs and the cumulative observed set $O_T$, where $T$ is the total number of decision rounds, are retained to evaluate tracing outcomes, investigation processes, and costs.

\subsection{Investigation Evaluation}
\label{sub:metrics}

AMLBench evaluates agents' performance in fund tracing across three dimensions: tracing results, investigation processes, and resource usage.
\ding{182}~We compare the transfers in the final submission against case annotations to measure the accuracy and completeness of the tracing results.
\ding{183}~We then examine transfers observed during querying alongside the final submission to determine which relevant transfers the agent retrieved and how many were included in the final result.
\ding{184}~We finally report decision rounds, tool call counts, and model token usage to compare the resources different agents require to obtain their tracing results.



\smallskip
\noindent\textbf{Tracing Performance.}
AMLBench compares the agent's final transfer set $\widehat F$ with the relevant transfer set $F\subseteq E$ defined by case annotations to evaluate tracing accuracy and completeness.
We compute transfer-level precision and recall as:
\[
P_{\mathrm{edge}}
=
\frac{|\widehat F\cap F|}{|\widehat F|},
\qquad
R_{\mathrm{edge}}
=
\frac{|\widehat F\cap F|}{|F|}.
\]
Here, $P_{\mathrm{edge}}$ measures the proportion of submitted transfers supported by case annotations, while $R_{\mathrm{edge}}$ measures the proportion of annotated relevant transfers recovered by the agent.
We report their harmonic mean as $F1_{\mathrm{edge}}$.

Since the initial clue set $S$ is provided at the start of the investigation, we also report $N_{\mathrm{new}}=|(\widehat F\cap F)\setminus S|$, the number of relevant transfers recovered beyond these clues.

In addition to transfer-level metrics, we evaluate recovery at the address level.
Let $V(A)$ denote the set of all sender and recipient addresses in a transfer set $A$.
Using $V(F)$ and $V(\widehat F)$ as the target and submitted address sets, respectively, we compute address-level precision, recall, and F1 analogously to complement the evaluation of transfer network recovery.

\smallskip
\noindent\textbf{Investigation Process.}
We use the cumulative observed set $O_T$ to compute observation coverage and, for $O_T\neq\varnothing$, the query hit rate:
\[
C_{\mathrm{obs}}
=
\frac{|O_T\cap F|}{|F|},
\qquad
H_{\mathrm{query}}
=
\frac{|O_T\cap F|}{|O_T|}.
\]
Observation coverage measures the proportion of annotated relevant transfers encountered in tool responses, while the query hit rate measures the proportion of observed transfers labeled as relevant to the case.
For runs with $O_T\cap F\neq\varnothing$, we further compute a retention rate to measure the proportion of these transfers included in the final submission:
\[
C_{\mathrm{ret}}
=
\frac{|\widehat F\cap O_T\cap F|}{|O_T\cap F|}.
\]

\smallskip
\noindent\textbf{Investigation Cost.}
We report the number of decision rounds $T$ and tool calls $N_{\mathrm{call}}$, together with the model-reported input and output token counts $N_{\mathrm{in}}$ and $N_{\mathrm{out}}$.
Total token usage is denoted by $N_{\mathrm{tok}}=N_{\mathrm{in}}+N_{\mathrm{out}}$.
Rounds and calls are reported separately because a single decision round may contain multiple tool calls.
These costs are presented alongside tracing metrics to compare the resources different agents require to obtain their tracing results~\cite{kapoor2025ai}.

\section{AMLAgent}
\label{sec-amlagent}
To support autonomous investigations on AMLBench, we implement AMLAgent, a tool-augmented agent for fund tracing.
AMLAgent uses an LLM to select investigation operations, identify relevant transfers, and generate the final submission.

\subsection{Agent Workflow}
\label{sec:agent_workflow}




AMLAgent organizes fund tracing as a multi-round decision process driven by query feedback, consistent with interactive LLM-agent workflows~\cite{yao2023react}.
Across cases, AMLAgent uses a common task input format, fund-tracing guidance, investigation operation interfaces, and execution workflow.
When comparing models on the same case, we keep these settings fixed and vary the underlying LLM.

Let $x$ denote the case task, which includes the case identifier, blockchain, investigation time window, and initial clues $S$.
We initialize the interaction history as $H_0=[x]$.
At round $t$, the model generates a decision based on the fund-tracing guidance $g$, operation descriptions $D_{\mathcal T}$, and current history $H_{t-1}$:
\[
a_t=\pi_{\theta}(g,D_{\mathcal T},H_{t-1}),
\]
where $\pi_{\theta}$ denotes the decision function implemented by the underlying LLM.
The decision $a_t$ may contain one or more investigation operations with their parameters, or a final submission.

When $a_t$ contains tool calls, AMLBench executes the corresponding operations and returns the results.
The response sequence for the round is denoted by $r_t=(r_{t,1},\ldots,r_{t,m_t})$, where $m_t$ is the number of tool calls in that round.
The effective history used in the next round is updated as
\[
H_t=\mathcal{C}\!\left(
H_{t-1}\oplus(a_t,r_t)
\right),
\]
where $\oplus$ denotes appending interaction content in chronological order, and $\mathcal{C}$ denotes the context management process described later.
The model selects subsequent investigation operations based on the updated history, adapting its actions to newly returned transaction information.

The investigation ends when the model issues no further tool calls and produces a valid structured result.
The final submission includes the transfer set $\widehat F$ identified as relevant to the case and a reason for stopping.
Each transfer is linked to a specific record in AMLBench through its unique identifier.






\subsection{Fund-Tracing Guidance}

AMLAgent uses shared guidance \mbox{$g=(g_{\mathrm{dir}},g_{\mathrm{adapt}},g_{\mathrm{verify}})$} to adapt general LLMs to AML fund tracing.
The components guide tracing direction, adaptation to fund-flow changes, and verification of candidate links and stopping decisions, respectively.
This guidance is provided to all underlying models through a common system prompt, leaving the model to select operations based on the current investigation information.

\smallskip
\noindent\textbf{Tracing Direction.}
We represent the initial clues as \mbox{$S=S^{\rightarrow}\cup S^{\leftarrow}$}, where $S^{\rightarrow}$ contains entry transfers for forward tracing and $S^{\leftarrow}$ contains downstream observations for tracing sources.
When no downstream observation is provided, $S^{\leftarrow}=\varnothing$.
The guidance $g_{\mathrm{dir}}$ directs the model to trace forward from entry transfers and backward from available downstream observations to check their fund-flow connections, while continuing to examine other branches supported by the records.


\smallskip
\noindent\textbf{Adaptive Tracing.}
The guidance $g_{\mathrm{adapt}}$ directs the model to adapt its queries as new information emerges.
The model uses observed counterparties and account activity to decide where to trace next, while tracking pagination and asset changes.
For potential swaps, it checks transfers within the same transaction to verify the asset change before continuing the trace.

\smallskip
\noindent\textbf{Fund-Flow Verification and Stopping.}
The guidance $g_{\mathrm{verify}}$ instructs the model to treat swap, splitting, and aggregation analysis results as candidate links requiring verification against specific transfer records.
It also instructs the model to submit observed transfers judged relevant to the case, requiring the final submission to satisfy $\widehat F\subseteq O_T.$
When the available records no longer support further tracing, or the remaining rounds are insufficient for verification, the model should consolidate and submit its findings.

\subsection{Context Management}




As an investigation progresses, accumulated interactions with investigation tools continually increase the context length~\cite{yang2024sweagent}.
AMLAgent therefore distinguishes the effective context $H_t$ used for subsequent decisions from the full investigation trajectory $\Gamma_t$ retained for process evaluation.

Following the definition in Agent Workflow~(\S\ref{sec:agent_workflow}), the context management function $\mathcal C$ transforms the history augmented with the current round's model outputs and tool responses into the effective context for the next round.
When no context processing is needed, $\mathcal C(H)=H$.
When the input token usage reported in the current round's model response reaches a preset threshold, AMLAgent completes the round's tool calls and appends their results before compacting the updated history to control subsequent model input size and support longer investigations.

Meanwhile, AMLAgent preserves the original, uncompressed investigation trajectory by appending new interactions.
With $\Gamma_0=[x]$, the update is \mbox{$\Gamma_t=\Gamma_{t-1}\oplus(a_t,r_t)$} after each round.
Here, $\Gamma_t$ retains model outputs, tool calls, and their responses across rounds.
Thus, even when the effective context used by the model is compacted, AMLBench can inspect the queries actually executed and their returned records through the original trajectory, and analyze investigation processes and resource usage using separately recorded token counts.

\section{Experimental Results}
 \label{sec-evalua}
We evaluate AMLAgent on AMLBench to assess how effectively LLM agents trace incident-related funds via autonomous investigation.
Since autonomous tracing involves multiple rounds of querying and decision-making, we examine both the investigation process and its outcomes to better understand agents' capabilities and limitations.
Specifically, we address the following research questions~(RQs):
\begin{packeditemize}
    \item \textbf{RQ1.} How accurately and completely can AMLAgent reconstruct incident-related transfer networks?
    \item \textbf{RQ2.} How many relevant transfers do agents retrieve, and how many of these are included in their final submissions?
    \item \textbf{RQ3.} What resources do autonomous investigations consume?
    \item \textbf{RQ4.} How do agents powered by different LLMs compare when investigating the same case?
\end{packeditemize}

\subsection{Experimental Setup}
\label{sec:exp_setup}




\noindent\textbf{Evaluation Settings.}
The cross-case evaluation covers all 24 cases in AMLBench, using GPT-6 Astra~\cite{openai_models} with one run per case and a 50-round limit.
We also compare different LLM backbones on Bybit, the largest case in AMLBench, to evaluate their performance on the same investigation task.

For the multi-model comparison, we select Bybit, the largest case in AMLBench with approximately 2.62 million transfer records, to examine model performance and resource usage at scale.
We evaluate GPT-6 Astra, GPT-5.6 Sol, GPT-5.6 Luna, and GPT-5.6 Terra~\cite{openai_models} with a 50-round limit. We conduct three runs per model.

\smallskip
\noindent\textbf{Agent Configuration.}
All experiments enable transaction record queries and fund-flow analysis, with the latter providing candidate swaps, splits, and aggregations for verification against specific transfer records.
All models receive the same fund-tracing guidance and tool parameter descriptions and may generate multiple tool calls in a single response.
Context compaction is triggered when the model-reported input token count for a round reaches \(200{,}000\).

\smallskip
\noindent\textbf{Result Reporting.}
We use the metrics defined in Section~\ref{sub:metrics} and report macro averages.
For the multi-model comparison, we report metric ranges across valid submissions and separately record submissions, round-limit exhaustion without submission, and execution failures.
Tracing metrics are computed from valid submissions, while completion statistics include all attempted runs.
Resource usage is calculated from run logs, with token counts covering both model decisions and context compaction.

\subsection{Fund-Tracing Performance (RQ1)}

We evaluate AMLAgent powered by GPT-6 Astra on 24 cases to assess its ability to recover case-related transfers and addresses from limited initial clues.
Table~\ref{tab:amlbench_24case_results} reports per-case results and macro-averaged metrics.
We make the following observations.

\ding{182}~\textbf{AMLAgent achieves good tracing accuracy and coverage on most cases.}
It achieves a transfer-level F1 above 70\% in 15 of the 24 cases, with a median F1 of 75.48\%.
Median transfer-level Precision and Recall are 95.10\% and 65.64\%, respectively, indicating that the agent generally submits annotation-supported transfers while recovering a substantial proportion of the target transfers.
\ding{183}~\textbf{AMLAgent discovers case-related addresses beyond the initial clues.}
It recovers additional case-related addresses not involved in the initial transfers in 23 of the 24 cases.
Across cases, these newly discovered addresses account for a median of 85.45\% of correctly recovered addresses, constituting a substantial part of the address recovery results.
Composable queries and fund-tracing guidance enable the agent to follow newly discovered addresses and examine their fund flows.
\ding{184}~\textbf{Large-scale cases remain challenging.}
Upbit, PlusToken, and Bybit each contain over 48,000 target transfers, compared with fewer than 4,000 in other cases. Although precision exceeds 85\% in all three, recall remains below 2\%. The agent retrieves fewer than 2.5\% of target transfers before stopping, leaving many fund-flow branches unexplored.
\ding{185}~\textbf{High recall does not always imply high precision.} In ATO and Bitpoint, AMLAgent achieves recall of 100.00\% and 94.23\%, respectively, but precision drops to 15.32\% and 7.56\%. In both cases, the agent submits many unannotated transfers alongside the relevant ones. This reveals a different limitation from large-scale cases, where incomplete exploration is the primary bottleneck.

\begin{table*}[t]
\centering
\caption{Investigation Evaluation on all 24 AMLBench cases using GPT-6 Astra.}
\label{tab:amlbench_24case_results}
\begingroup
\fontsize{8.5}{10.2}\selectfont
\setlength{\tabcolsep}{3.3pt}
\renewcommand{\arraystretch}{1.02}
\begin{tabular*}{\textwidth}{@{\extracolsep{\fill}}c|rrrr|rrr|rrr|rrr|c@{}}
\toprule
\multicolumn{1}{c}{\multirow{2}{*}{\textbf{Case}}}
& \multicolumn{4}{c}{\textbf{Transfer counts}}
& \multicolumn{3}{c}{\textbf{Transfer-level} (\%)}
& \multicolumn{3}{c}{\textbf{Address-level} (\%)}
& \multicolumn{3}{c}{\textbf{Process} (\%)}
& \textbf{Tokens} {(M)} \\
\cmidrule(lr){2-5}
\cmidrule(lr){6-8}
\cmidrule(lr){9-11}
\cmidrule(lr){12-14}
\cmidrule(l){15-15}
& $|S|$ & $|F|$ & $|\widehat F|$ & $N_{\mathrm{new}}$
& P & R & F1
& P & R & F1
& $C_{\mathrm{obs}}$ & $H_{\mathrm{query}}$ & $C_{\mathrm{ret}}$
& $N_{\mathrm{tok}}$ \\
\midrule
AscendEX & 93 & 1,389 & 1,010 & 881 & 96.44 & 70.12 & 81.20 & 97.80 & 53.94 & 69.53 & 93.95 & 41.43 & 74.64 & 2.74 \\
ATO & 1 & 51 & 333 & 51 & 15.32 & 100.00 & 26.56 & 0.70 & 100.00 & 1.40 & 100.00 & 15.27 & 100.00 & 0.95 \\
BadgerDAO & 1 & 189 & 6 & 0 & 16.67 & 0.53 & 1.03 & 28.57 & 7.41 & 11.76 & 100.00 & 75.00 & 0.53 & 0.38 \\
Bancor & 2 & 167 & 107 & 104 & 99.07 & 63.47 & 77.37 & 85.71 & 54.55 & 66.67 & 100.00 & 85.64 & 63.47 & 0.14 \\
BitMart & 86 & 700 & 647 & 479 & 87.33 & 80.71 & 83.89 & 47.46 & 90.32 & 62.22 & 100.00 & 75.84 & 80.71 & 1.58 \\
Bitpoint & 2 & 52 & 648 & 47 & 7.56 & 94.23 & 14.00 & 1.16 & 87.50 & 2.28 & 100.00 & 7.90 & 94.23 & 0.59 \\
Bybit & 5 & 199,518 & 2,085 & 1,959 & 94.20 & 0.98 & 1.95 & 99.20 & 2.15 & 4.21 & 2.19 & 77.88 & 44.92 & 7.13 \\
Cashio & 6 & 30 & 25 & 18 & 96.00 & 80.00 & 87.27 & 85.71 & 75.00 & 80.00 & 100.00 & 53.57 & 80.00 & 0.10 \\
Cream Finance & 3 & 391 & 333 & 251 & 76.28 & 64.96 & 70.17 & 89.83 & 74.65 & 81.54 & 100.00 & 50.65 & 64.96 & 1.64 \\
DEGO \& Cocos & 2 & 20 & 32 & 18 & 62.50 & 100.00 & 76.92 & 50.00 & 100.00 & 66.67 & 100.00 & 62.50 & 100.00 & 0.09 \\
EasyFi & 20 & 1,119 & 800 & 718 & 92.25 & 65.95 & 76.92 & 80.00 & 41.03 & 54.24 & 96.43 & 52.92 & 68.40 & 2.76 \\
Fake Metadium & 2 & 478 & 411 & 405 & 99.03 & 85.15 & 91.56 & 98.85 & 90.53 & 94.51 & 97.49 & 85.19 & 87.34 & 1.62 \\
Harvest Finance & 2 & 1,419 & 990 & 960 & 97.17 & 67.79 & 79.87 & 86.21 & 30.49 & 45.05 & 83.23 & 64.12 & 81.46 & 1.98 \\
LCX & 11 & 129 & 112 & 101 & 100.00 & 86.82 & 92.95 & 100.00 & 100.00 & 100.00 & 100.00 & 88.97 & 86.82 & 0.16 \\
Lendf.Me & 2 & 529 & 338 & 334 & 99.41 & 63.52 & 77.51 & 91.67 & 95.65 & 93.62 & 100.00 & 16.96 & 63.52 & 2.45 \\
Liquid Exchange & 54 & 3,192 & 1,881 & 1,824 & 99.84 & 58.83 & 74.04 & 96.59 & 50.30 & 66.15 & 98.87 & 92.77 & 59.51 & 3.43 \\
Nexus Mutual & 10 & 706 & 152 & 119 & 84.87 & 18.27 & 30.07 & 71.05 & 52.94 & 60.67 & 88.39 & 88.51 & 20.67 & 1.51 \\
PlusToken & 3 & 48,074 & 903 & 900 & 100.00 & 1.88 & 3.69 & 100.00 & 2.60 & 5.08 & 2.49 & 91.16 & 75.50 & 3.94 \\
Poly Network & 2 & 224 & 101 & 76 & 77.23 & 34.82 & 48.00 & 52.17 & 32.43 & 40.00 & 100.00 & 8.67 & 34.82 & 2.04 \\
SpankChain & 2 & 32 & 53 & 29 & 56.60 & 93.75 & 70.59 & 60.00 & 100.00 & 75.00 & 100.00 & 33.68 & 93.75 & 0.24 \\
Upbit & 2 & 431,719 & 1,800 & 1,531 & 85.17 & 0.36 & 0.71 & 81.46 & 2.78 & 5.38 & 0.62 & 83.91 & 56.86 & 4.36 \\
Uranium Finance & 13 & 45 & 45 & 31 & 97.78 & 97.78 & 97.78 & 90.91 & 100.00 & 95.24 & 100.00 & 90.00 & 97.78 & 0.09 \\
Vulcan Forged & 9 & 1,090 & 712 & 703 & 100.00 & 65.32 & 79.02 & 100.00 & 78.79 & 88.14 & 100.00 & 83.91 & 65.32 & 1.16 \\
Warp Finance & 2 & 276 & 107 & 104 & 99.07 & 38.41 & 55.35 & 96.55 & 68.29 & 80.00 & 69.93 & 70.96 & 54.92 & 0.48 \\

\midrule

\multicolumn{1}{c|}{\textbf{Macro}} & \textemdash & \textemdash & \textemdash & \textemdash & 80.82 & 59.74 & 58.27 & 74.65 & 62.14 & 56.22 & 84.73 & 62.39 & 68.76 & 1.73 \\
\bottomrule
\end{tabular*}
\endgroup
\end{table*}

\subsection{Investigation Process Analysis (RQ2)}

\begin{figure}[t]
    \centering
    \includegraphics[width=0.9\linewidth]{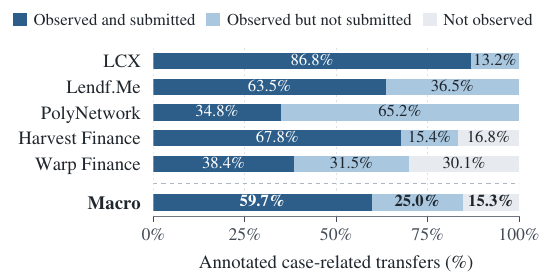}
    \caption{Transfer observation and submission in representative cases.
    Macro denotes the mean value across all 24 cases.}
        \vspace{-0.1in}
    \label{fig:investigation_process}
\end{figure}

\begin{figure}[t]
    \centering
    \includegraphics[width=0.88\linewidth]{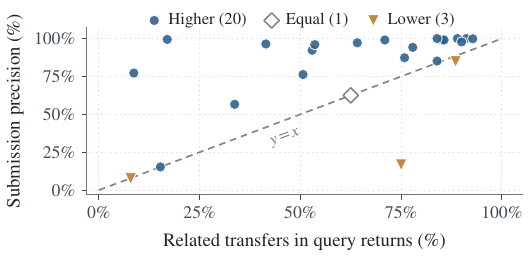}
      \vspace{-0.1in}
    \caption{Transfer selection across AMLBench cases.}
    \label{fig:transfer_selection}
\end{figure}


\begin{figure}[!]
    \centering
    \includegraphics[width=0.85\linewidth]{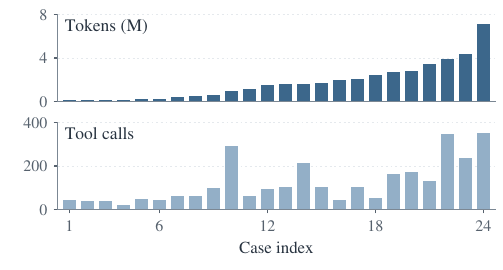}
     \vspace{-0.1in}
    \caption{Token usage (top) and tool calls (bottom) with the same case order, from lowest to highest token usage.}
    \label{fig:cost_tool}
        \vspace{-0.1in}
\end{figure}


We examine query responses alongside final submissions to analyze how AMLAgent retrieves and selects case-related transfers.
Table~\ref{tab:amlbench_24case_results} reports process metrics for all cases.
Figures~\ref{fig:investigation_process} and~\ref{fig:transfer_selection} show transfer observation and submission outcomes and changes in the proportion of case-related transfers after selection, respectively. We make the following observations.

\ding{182}~\textbf{AMLAgent retrieves most case-related transfers in most cases.} (Table~\ref{tab:amlbench_24case_results})
Macro-average query coverage is 84.73\%, with 18 cases reaching at least 90\% coverage, including 14 with full coverage of annotated case-related transfers.
Composable queries support continued retrieval from newly discovered accounts and transactions.
\ding{183}~\textbf{Final selection improves submission precision in most cases.} (Figures~\ref{fig:investigation_process}~and~\ref{fig:transfer_selection}) 
The proportion of case-related transfers increases after selection in 20 cases.
Across all cases, the median proportions in query responses and final submissions are 72.98\% and 95.10\%, respectively.
The model uses fund-flow connections to select records from account activity for final submission.
\ding{184}~\textbf{Some observed transfers are not submitted because of zero values or unclear fund-flow connections.}
The macro-average retention rate is 68.76\%.
In Lendf.Me, 193 annotated case-related transfers were observed but not submitted, and 192 of them have zero amounts.
In Vulcan Forged, the agent excluded some outflows because it could not confirm their connection to the initial funds after mixing.

\subsection{Investigation Cost (RQ3)}



We analyze AMLAgent's resource usage across the 24 cases.
Table~\ref{tab:amlbench_24case_results} reports per-case token usage, while Figure~\ref{fig:cost_tool} shows the relationship between tool call counts and token consumption. Detailed per-case decision rounds and tool call counts are reported in Appendix Table~\ref{tab:investigation_cost}. We make the following observations.

\ding{182}~\textbf{Investigation costs vary substantially across cases.}
Token usage ranges from 0.09M to 7.13M per case, with a mean of 1.73M.
Upbit, PlusToken, and Bybit consume 4.36M, 3.94M, and 7.13M tokens, respectively, making them the three cases with the highest token usage.
These results show that actual resource requirements vary considerably across cases under the same tool configuration and investigation round limit.
\ding{183}~\textbf{Fewer tool calls do not necessarily imply lower token usage.}
ATO makes 292 tool calls and consumes 0.95M tokens, whereas Lendf.Me makes only 50 calls but consumes 2.45M tokens.
The two investigations observe 334 and 3,120 unique transfers, respectively, showing that tool call counts alone do not fully reflect the number of records retrieved.

\subsection{Multi-Model Comparison (RQ4)}
\begin{table}[b]
\centering
\caption{Multi-model comparison on Bybit.}
\label{tab:bybit_model_comparison}
\begingroup
\fontsize{8.5}{10.2}\selectfont
\renewcommand{\arraystretch}{1.08}
\begin{tabular*}{0.47\textwidth}{@{\extracolsep{\fill}}l|rr|rr|c@{}}
\toprule
\multicolumn{1}{c}{\multirow{2}{*}{\textbf{Model}}}
& \multicolumn{2}{c}{\textbf{Precision} }
& \multicolumn{2}{c}{\textbf{Process} }
& \multirow{2}{*}{\shortstack{\textbf{Tokens}\\(M)}} \\
\cmidrule(lr){2-3}\cmidrule(lr){4-5}
& $P_{\mathrm{edge}}$ & $P_{\mathrm{addr}}$
& $C_{\mathrm{obs}}$ & $C_{\mathrm{ret}}$ & \\
\midrule
GPT-6 Astra   & 91.64 & 98.83  & 1.95 & 52.24 & 4.89 \\
GPT-5.6 Sol   & 93.46 & 100.00 & 2.19 & 4.58  & 2.74 \\
GPT-5.6 Luna  & 95.00 & 100.00 & 1.82 & 2.37  & 3.53 \\
GPT-5.6 Terra & 87.41 & 95.71  & 0.89 & 3.67  & 1.38 \\
\bottomrule
\end{tabular*}
\endgroup
\end{table}





We compare AMLAgent powered by four LLMs on Bybit, keeping the initial clues, fund-tracing guidance, tool configuration, and round limit fixed.
Table~\ref{tab:bybit_model_comparison} reports mean results, and Multi-model evaluation results on more AMLBench cases are reported in Appendix~\ref{app:multi}. We make the following observations.

\ding{182}~\textbf{All models achieve high submission precision.}
Transfer-level precision ranges from 87.41\% to 95.00\%, while address-level precision ranges from 95.71\% to 100.00\%.
These results indicate that most submitted transfers and addresses are supported by the case annotations.
\ding{183}~\textbf{Models differ substantially in their retention of observed relevant transfers.}
Sol achieves the highest query coverage of 2.19\%, compared with 1.95\% for Astra.
However, Astra retains 52.24\% of the observed relevant transfers, while all other models retain less than 5\%.
Thus, even with similar query coverage, models can differ substantially in the proportion of relevant transfers included in their final submissions, highlighting the importance of result consolidation and transfer selection.
\ding{184}~\textbf{Token usage varies substantially under the same round limit.}
Mean token usage ranges from 1.38M for Terra to 4.89M for Astra, a ratio of approximately 3.5 to 1.
Astra retains a larger proportion of observed relevant transfers while consuming more tokens, highlighting the need to consider both investigation output and resource usage when comparing models.

\subsection{Case Study}


In the Vulcan Forged case, AMLAgent starts from eight known PYR incoming transfers and one downstream ETH transfer. It queries the transaction histories of related accounts and follows newly discovered addresses to trace further fund flows. For accounts with extensive transaction histories, the Agent continues retrieving additional records. When asset conversions occur, it examines transfers within the same transaction to verify connections between different assets. Through this process, AMLAgent observes 1,299 transfers and ultimately identifies and submits 712 transfers, all supported by the case ground truth. Among them, 703 are newly recovered transfers beyond the initial clues.

\begin{figure}[H]
    \centering
    \includegraphics[width=0.9\linewidth]{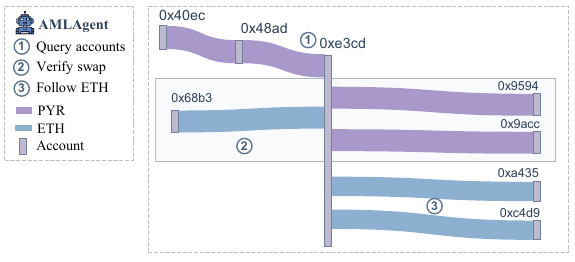}
    \caption{Case study of AMLAgent in Vulcan Forged.}
    \label{fig:case_study}
\end{figure}

Figure~\ref{fig:case_study} illustrates a local fund flow involving PYR transfers, asset conversion, and subsequent ETH distribution. \ding{182}~AMLAgent queries account transaction histories and identifies a PYR transfer from 0x48ad to 0xe3cd. \ding{183}~By examining transfers within the same transaction, it verifies the PYR transfers from 0xe3cd to 0x9594 and 0x9acc, as well as the ETH received from 0x68b3, revealing the asset conversion relationship highlighted in the figure.\ding{184}~On the ETH side, the Agent further investigates recipient accounts such as 0xa435 and connects the known downstream transfer involving 0xc4d9 to the preceding PYR flow. This process demonstrates AMLAgent's ability to integrate account-level queries with transaction-level verification, enabling cross-asset fund tracing and continued investigation after asset conversion.

\section{Discussion}
\label{sec-discu}

Our design contributes to future AML investigation in two ways. 

\subsection{Towards Real-Time Detection and Tracing}

Our current evaluation focuses on historical transaction records that remain fixed throughout an investigation. In practice, however, illicit funds may continue moving while an investigation is underway. Supporting real-time tracing would therefore require AMLAgent to incorporate newly available evidence and update its findings as the investigation progresses.

AMLAgent's iterative workflow provides a natural basis for this extension. Existing AML monitoring systems, including those used in traditional banking, provide initial alerts when suspicious activity can be linked to on-chain transactions. AMLAgent traces the funds and preserves its findings as new blocks arrive. When additional transfers involve addresses under investigation, the agent can continue tracing the affected branches without repeating earlier queries. This would complement existing detection systems by extending initial alerts into ongoing fund-flow investigations.

AMLBench further provides a basis for evaluating real-time investigations at scale. Its timestamped transaction records and annotated fund flows could support chronological replay of real incidents, allowing agents to investigate transactions as they become available. This would enable systematic evaluation of how quickly agents identify relevant fund movements and how effectively they reconstruct evolving fund flows within a response deadline.

\subsection{Strengthening Autonomous Fund Tracing}

Our results highlight the importance of managing investigation progress. In the largest cases, AMLAgent achieves high submission precision but observes only a small fraction of relevant transfers. Maintaining an explicit record of unexplored branches and unfinished queries could help the agent prioritize further exploration and avoid premature stopping. This would provide a clearer basis for deciding when an investigation is sufficiently complete.

Evidence management is another important aspect. Our Bybit experiments show that similar levels of evidence retrieval can lead to substantially different final results. A persistent evidence record could help AMLAgent preserve relevant transfers and their fund-flow connections throughout long investigations, even when conversation history is compacted. However, not all omitted transfers represent overlooked evidence. In Lendf.Me and Vulcan Forged, some transfers were excluded because of zero amounts or uncertain fund-flow connections. Recording these decisions would help distinguish deliberate exclusions from unintended omissions.

Resource management also affects investigation quality. Our experiments show that fewer tool calls do not necessarily lead to lower token consumption, as individual queries may return substantially different amounts of data. More focused queries reduce processing overhead but may also miss relevant fund flows. Effective investigation strategies should therefore prioritize queries that provide useful evidence while making efficient use of available resources.

\section{Conclusion}

We presented \textit{AMLBench}, a benchmark for evaluating autonomous fund tracing from real-world incidents, and \textit{AMLAgent}, a reference agent investigating the cases from limited clues. We experimentally demonstrate the effectiveness of AMLAgent. We provide a reproducible foundation for  advancing agentic on-chain AML research.




\bibliographystyle{ACM-Reference-Format}
\bibliography{reff}

\appendix

\section*{Appendix}

\section{Extended Comparison with Prior Work (\S\ref{sec-rw})}
\label{app:related_work}

\begin{table}[!t]
    \centering
    \fontsize{7.5}{9}\selectfont
    \setlength{\tabcolsep}{1pt}
    \renewcommand{\arraystretch}{1.1}
    \caption{Detailed statistics of the 24 cases in AMLBench.}
    \label{tab:case_statistics}
    \begin{tabular*}{\columnwidth}{
        @{\extracolsep{\fill}}ccccccc@{}
    }
        \toprule
        \multicolumn{1}{c}{\textbf{Case}} &  \multicolumn{1}{c}{\textbf{Year}} 
        & \shortstack{\textbf{Transfers}}
        & \textbf{Addresses}
        & \shortstack{\textbf{Laundering}\\\textbf{transfers}}
        & \shortstack{\textbf{Data span}\\(days)}
        & \shortstack{\textbf{Initial}\\\textbf{clues}} \\
        \midrule
        Bancor
        & 2018 & 195 & 27 & 167 & 1,273.81 & 2 \\
        Fake Metadium
        & 2018 & 572 & 264 & 478 & 1,636.70 & 2 \\
        SpankChain
        & 2018 & 95 & 28 & 32 & 20.13 & 2 \\
        Bitpoint
        & 2019 & 7,439 & 7,357 & 52 & 121.69 & 2 \\
        PlusToken
        & 2019 & 64,747 & 38,327 & 48,074 & 871.12 & 3 \\
        Upbit
        & 2019 & 2,303,765 & 577,994 & 431,719 & 2,309.78 & 2 \\
        Harvest Finance
        & 2020 & 2,628 & 113 & 1,419 & 834.06 & 2 \\
        Lendf.Me
        & 2020 & 3,120 & 78 & 529 & 597.60 & 2 \\
        Nexus Mutual
        & 2020 & 959 & 134 & 706 & 1,545.21 & 10 \\
        Warp Finance
        & 2020 & 408 & 78 & 276 & 265.08 & 2 \\
        AscendEX
        & 2021 & 28,991 & 6,713 & 1,389 & 675.00 & 93 \\
        BadgerDAO
        & 2021 & 252 & 48 & 189 & 252.00 & 1 \\
        BitMart
        & 2021 & 923 & 133 & 700 & 319.19 & 86 \\
        Cream Finance
        & 2021 & 775 & 213 & 391 & 232.54 & 3 \\
        EasyFi
        & 2021 & 8,811 & 1,378 & 1,119 & 1,937.68 & 20 \\
        Liquid Exchange
        & 2021 & 6,174 & 381 & 3,192 & 1,162.46 & 54 \\
        Poly Network
        & 2021 & 2,584 & 1,391 & 224 & 374.35 & 2 \\
        Uranium Finance
        & 2021 & 50 & 15 & 45 & 441.06 & 13 \\
        Vulcan Forged
        & 2021 & 3,911 & 564 & 1,090 & 364.60 & 9 \\
        ATO 
        & 2022 & 334 & 285 & 51 & 0.31 & 1 \\
        Cashio
        & 2022 & 56 & 25 & 30 & 55.24 & 6 \\
        DEGO \& Cocos
        & 2022 & 32 & 18 & 20 & 0.60 & 2 \\
        LCX
        & 2022 & 145 & 27 & 129 & 31.55 & 11 \\
        Bybit
        & 2025 & 2,616,736 & 58,790 & 199,518 & 14.00 & 5 \\
        \midrule
        \multicolumn{1}{c}{\textbf{Total}}
        & -- & 5,053,702 & 694,381  & 691,539 & -- & 335 \\
        \bottomrule
    \end{tabular*}
\end{table}

We organize this section around three questions. We first examine what blockchain forensics currently automates and then consider why laundering infrastructure defeats static tracing. We finally turn to how progress on LLM agents is measured.  



\subsection{Automating Blockchain Forensics}

Since transaction histories became publicly auditable~\cite{nakamoto2008bitcoin}, forensics research has automated attribution first. Clustering heuristics resolve pseudonymous addresses into entities on Bitcoin~\cite{meiklejohn2013fistful,androulaki2013evaluating,ron2013quantitative}, and later work adapted them to Ethereum's account model~\cite{victor2020address}. Analysis platforms such as BlockSci and GraphSense, followed by systems including BITSQL, industrialized this layer by turning raw ledgers into queryable infrastructure over which heuristics and measurements can be reproduced at scale~\cite{kalodner2020blocksci,haslhofer2021graphsense,mun2022bitsql}.

Learning-based detection built on this infrastructure and extended coverage. The Elliptic dataset established graph-learning baselines for illicit-transaction classification~\cite{weber2019anti}, while Elliptic2 raised the unit of analysis to laundering subgraphs~\cite{bellei2024shape}. BABD-13 further supplied labeled Bitcoin address behaviors at scale~\cite{xiang2024babd}. Subsequent detectors have targeted illicit accounts and phishing~\cite{farrugia2020detection,wu2020phishers}, as well as replicated scam campaigns and Ponzi contracts~\cite{phillips2020tracing,chen2018detecting,bartoletti2020dissecting}.

Each of these systems automates one sub-question with a fixed input and a fixed output: an address in, a label out; a subgraph in, a score out. What none of them automates is the step that consumes analyst time in practice, assembling those sub-answers into a coherent, evidenced account of a specific incident. A benchmark for forensic agents must therefore pose end-to-end investigations rather than component predictions.

\subsection{Laundering Against Static Tracing}

The services that launder stolen value can be read as attacks on specific tracing assumptions. Mixers attack the link between inputs and outputs, from early centralized tools~\cite{moser2013inquiry} to zero-knowledge designs, though measurement studies show that operational mistakes and detectable structure often restore partial traceability~\cite{pakki2021everything,wu2021detecting,beres2021blockchain}. Privacy-focused ledgers weaken graph visibility by design, yet empirical analyses of CryptoNote chains and Zcash reach a similar conclusion: anonymity degrades under realistic usage~\cite{moser2018empirical,yu2019new,kappos2018empirical}.

Other services attack the analysis scope rather than the link structure. Bridges and exchange routes move value beyond any single ledger, motivating cross-ledger correlation~\cite{yousaf2019tracing} and creating interoperability-specific risks~\cite{augusto2024sok,wang2023exploring}. More recent evasion undermines the transfer abstraction itself. Execution can be separated from the principal and beneficiary while value migrates through contract state~\cite{cao2026peb}, and control over transaction ordering allows sanctioned flows to bypass address-based enforcement~\cite{wu2026ordering}. Compliance research responds with privacy-preserving AML checks~\cite{femiak2025zero}, but enforcement and evasion continue to co-evolve.

\begin{figure}[!]
    \centering
    \includegraphics[width=\columnwidth]
        {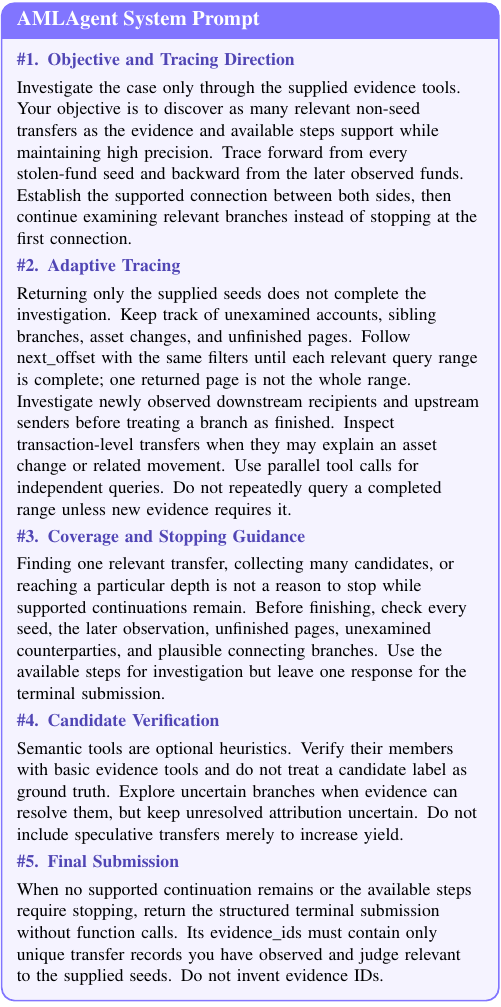}
        \vspace{-0.15in}
    \caption{System prompt of AMLAgent.}
    \label{fig:amlagent_system_prompt}
\end{figure}

\begin{table}[t]
\centering
\caption{Per-case investigation costs using GPT-6 Astra.}
\label{tab:investigation_cost}
\begingroup
\fontsize{7.5}{9}\selectfont
\setlength{\tabcolsep}{4pt}
\renewcommand{\arraystretch}{1}
\begin{tabular*}{\linewidth}{@{\extracolsep{\fill}}cccc@{}}
\toprule
\multicolumn{1}{c}{\textbf{Case}} & \textbf{Rounds} ($T$) & \textbf{Calls} ($N_{\mathrm{call}}$) & \textbf{Tokens/M} ($N_{\mathrm{tok}}$ ) \\ 
\midrule
AscendEX & 27 & 162 & 2.74 \\
ATO & 16 & 292 & 0.95 \\
BadgerDAO & 11 & 61 & 0.38 \\
Bancor & 5 & 19 & 0.14 \\
BitMart & 12 & 105 & 1.58 \\
Bitpoint & 10 & 100 & 0.59 \\
Bybit & 48 & 354 & 7.13 \\
Cashio & 7 & 40 & 0.10 \\
Cream Finance & 18 & 104 & 1.64 \\
DEGO \& Cocos & 7 & 43 & 0.09 \\
EasyFi & 25 & 173 & 2.76 \\
Fake Metadium & 17 & 213 & 1.62 \\
Harvest Finance & 17 & 45 & 1.98 \\
LCX & 6 & 48 & 0.16 \\
Lendf.Me & 19 & 50 & 2.45 \\
Liquid Exchange & 24 & 132 & 3.43 \\
Nexus Mutual & 13 & 93 & 1.51 \\
PlusToken & 46 & 349 & 3.94 \\
Poly Network & 21 & 102 & 2.04 \\
SpankChain & 9 & 42 & 0.24 \\
Upbit & 34 & 238 & 4.36 \\
Uranium Finance & 6 & 37 & 0.09 \\
Vulcan Forged & 13 & 61 & 1.16 \\
Warp Finance & 11 & 63 & 0.48 \\
\midrule
\textbf{Mean} & 17.58 & 121.92 & 1.73 \\
\bottomrule
\end{tabular*}
\endgroup
\end{table}

Real incidents compose several of these techniques in sequence, as documented by attack systematizations and incident repositories~\cite{zhou2023sok,defihacklabs,chainalysis2025midyear}, and the composition differs case by case. This variability is the reason fixed pipelines underperform, and it sets the second requirement: benchmark cases must be drawn from real incidents that combine heterogeneous services, not from synthetic or single-mechanism traffic.


\subsection{Measuring Progress on LLM Agents}

Benchmarks have been the instrument that made agent progress legible. General suites measure planning and tool use across operating-system and web environments, with others targeting assistant tasks or user interaction~\cite{liu2024agentbench,zhou2024webarena,mialon2023gaia,yao2024tau}. The most diagnostic results, however, come from domains where success is checked mechanically rather than judged by plausibility: SWE-bench scores patches against repository test suites~\cite{jimenez2024swebench}, and Cybench grades capture-the-flag performance at the subtask level~\cite{zhang2025cybench}.

The underlying techniques combine interleaved reasoning and action with external tool use, while self-correction allows agents to revise unsuccessful trajectories~\cite{yao2023react,schick2023toolformer,shinn2023reflexion}. These techniques have meanwhile reached blockchain security. LLMs are used to audit smart contracts and flag anomalous transactions~\cite{sun2024gptscan}. They have also been applied to interpreting cryptocurrency transaction graphs~\cite{lei2025large}. Fully agentic systems go further by collecting and reasoning over multi-chain evidence to conduct blockchain investigations~\cite{yu2026locard}, and related systems reconstruct exploit lifecycles with executable proofs of concept~\cite{xiao2026prompt,su2026transactions}.

The measurement side has not followed. These systems are validated on isolated mechanisms, such as cross-chain transfer matching, or on curated exploit code~\cite{xiao2026prompt}, each under study-specific criteria that do not transfer across systems. Forensics therefore still lacks what gave SWE-bench and Cybench their diagnostic power, and supplying it is the third requirement: ground truth that is verifiable on chain and shared across systems.

\begin{table}[t]
\centering
\small
\setlength{\tabcolsep}{3.5pt}
\renewcommand{\arraystretch}{1}
\caption{Comparison with existing solutions.}
\label{tab}
\vspace{-0.1in}
\begin{threeparttable}
\begin{tabular}{r ccccccc}
\toprule
\multicolumn{1}{c}{\textbf{Work}}
& \textbf{BC} & \textbf{RW}
& \textbf{GT} & \textbf{LT} & \textbf{EV}
& \textbf{AG} & \textbf{MC} \\
\midrule
Elliptic~\cite{weber2019anti}
& \fullcircle & \halfcircle
& \emptycircle & \emptycircle & \emptycircle
& \emptycircle & \emptycircle \\
Elliptic2~\cite{bellei2024shape}
& \fullcircle & \halfcircle
& \emptycircle & \halfcircle & \emptycircle
& \emptycircle & \emptycircle \\
BABD-13~\cite{xiang2024babd}
& \fullcircle & \halfcircle
& \emptycircle & \halfcircle & \halfcircle
& \emptycircle & \emptycircle \\
GAIA~\cite{mialon2023gaia}
& \emptycircle & \halfcircle
& \emptycircle & \emptycircle & \emptycircle
& \fullcircle & \emptycircle \\
Cybench~\cite{zhang2025cybench}
& \emptycircle & \halfcircle
& \emptycircle & \emptycircle & \halfcircle
& \fullcircle & \emptycircle \\
RiskTagger~\cite{lin2025risktagger}
& \fullcircle & \fullcircle
& \emptycircle & \halfcircle & \fullcircle
& \halfcircle & \fullcircle \\
Thor25~\cite{yu2026locard}
& \fullcircle & \halfcircle
& \halfcircle & \emptycircle & \halfcircle
& \halfcircle & \emptycircle \\
\midrule
\textbf{Our work}
& \fullcircle & \fullcircle
& \fullcircle & \fullcircle & \fullcircle
& \fullcircle & \fullcircle \\
\bottomrule
\end{tabular}
\begin{tablenotes}[flushleft]
\footnotesize
\item \textbf{BC}: blockchain-oriented setting; \textbf{RW}: real-world incident cases; \textbf{GT}: ground-truth fund traces; LT: laundering technique annotations; \textbf{EV}: auditable on-chain evidence; \textbf{AG}: agentic evaluation; \textbf{MC}: analysis that assembles evidence from multiple cases. 
\fullcircle: full support; \halfcircle: partial support; \emptycircle: no support.
\end{tablenotes}
\end{threeparttable}
\vspace{-0.15in}
\end{table}

\subsection{Research Gap}

The requirements above have each appeared in prior work, but never together. Detection datasets provide scale without investigations, whereas incident repositories provide realism without fund traces. Agent benchmarks offer verifiable grading without forensic content, while agentic forensic systems support investigations without common evaluation.

Table~\ref{tab} summarizes this landscape along the criteria a forensic benchmark must meet. AMLBench is designed to sit at the intersection. It organizes real incidents under a shared evidence schema containing ground-truth traces and technique annotations, with auditable on-chain evidence supporting the resulting records. AMLAgent provides a reference point for evaluating agents against the benchmark, including comparison across cases.

\begin{table*}[t]
    \centering
    \footnotesize
    \setlength{\tabcolsep}{2.5pt}
    \renewcommand{\arraystretch}{1.12}
    \caption{Mean performance on three cases across four models.}
    \label{tab:small_case_model_comparison}

    \begin{tabular*}{\textwidth}{
        @{\extracolsep{\fill}}
        c|l|rrrr|rrr|rrr|rrr|c
        @{}
    }
        \midrule
        
         \multicolumn{1}{c}{\multirow{2}{*}{\textbf{Case}}}
        & \multicolumn{1}{c}{\multirow{2}{*}{\textbf{Model}}}
        & \multicolumn{4}{c}{\textbf{Transfer counts}}
        & \multicolumn{3}{c}{\textbf{Transfer-level} (\%)}
        & \multicolumn{3}{c}{\textbf{Address-level} (\%)}
        & \multicolumn{3}{c}{\textbf{Process} (\%)}
        & \textbf{Tokens} (M) \\
        
        \cmidrule{3-6}\cmidrule{7-16}\cmidrule{3-16}
        
       \multicolumn{1}{c}{} &  & $|S|$ & $|F|$ & $|\widehat F|$ & $N_{\mathrm{new}}$
        & P & R & F1
        & P & R & F1
        & $C_{\mathrm{obs}}$ & $H_{\mathrm{query}}$ & $C_{\mathrm{ret}}$
        & $N_{\mathrm{tok}}$ \\
        
        \midrule

        \multirow{4}{*}{Cashio}
        & GPT-6 Astra
        & 6 & 30 & 25.67 & 18.00
        & 93.54 & 80.00 & 86.23
        & 78.57 & 75.00 & 76.67
        & 100.00 & 53.57 & 80.00
        & 0.13 \\
        & GPT-5.6 Sol
        & 6 & 30 & 17.00 & 10.33
        & 96.28 & 54.44 & 69.41
        & 91.53 & 83.33 & 86.61
        & 100.00 & 53.57 & 54.44
        & 0.18 \\
        & GPT-5.6 Luna
        & 6 & 30 & 15.00 & 9.00
        & 100.00 & 50.00 & 66.67
        & 100.00 & 75.00 & 85.71
        & 97.78 & 53.01 & 51.15
        & 0.14 \\
        & GPT-5.6 Terra
        & 6 & 30 & 15.00 & 9.00
        & 100.00 & 50.00 & 66.67
        & 100.00 & 75.00 & 85.71
        & 96.67 & 52.73 & 51.72
        & 0.05 \\
        
        \midrule

        \multirow{4}{*}{Uranium Finance}
        & GPT-6 Astra
        & 13 & 45 & 45.00 & 31.00
        & 97.78 & 97.78 & 97.78
        & 90.91 & 100.00 & 95.24
        & 100.00 & 90.00 & 97.78
        & 0.09 \\
        & GPT-5.6 Sol
        & 13 & 45 & 44.67 & 31.00
        & 98.52 & 97.78 & 98.14
        & 93.94 & 100.00 & 96.83
        & 100.00 & 90.00 & 97.78
        & 0.08 \\
        & GPT-5.6 Luna
        & 13 & 45 & 45.00 & 31.00
        & 97.78 & 97.78 & 97.78
        & 90.91 & 100.00 & 95.24
        & 100.00 & 90.00 & 97.78
        & 0.18 \\
        & GPT-5.6 Terra
        & 13 & 45 & 44.00 & 31.00
        & 100.00 & 97.78 & 98.88
        & 100.00 & 100.00 & 100.00
        & 100.00 & 90.00 & 97.78
        & 0.11 \\
        
         \midrule

        \multirow{4}{*}{SpankChain}
        & GPT-6 Astra
        & 2 & 32 & 47.00 & 29.00
        & 66.31 & 93.75 & 76.91
        & 60.00 & 100.00 & 75.00
        & 100.00 & 33.68 & 93.75
        & 0.20 \\
        & GPT-5.6 Sol
        & 2 & 32 & 60.00 & 29.67
        & 52.15 & 95.83 & 67.18
        & 58.18 & 100.00 & 73.53
        & 100.00 & 33.68 & 95.83
        & 0.23 \\
        & GPT-5.6 Luna
        & 2 & 32 & 35.33 & 24.33
        & 72.78 & 79.17 & 73.94
        & 47.81 & 94.44 & 62.42
        & 100.00 & 33.68 & 79.17
        & 0.28 \\
        & GPT-5.6 Terra
        & 2 & 32 & 34.00 & 24.33
        & 75.71 & 79.17 & 75.62
        & 49.63 & 94.44 & 63.89
        & 100.00 & 33.68 & 79.17
        & 0.27 \\
         \midrule
    \end{tabular*}
\end{table*}

\section{LLM Prompt Templates (\S\ref{sec-amlagent})}

We present the system and task prompts used by AMLAgent to guide and initialize investigations.

\subsection{System Prompt}


Figure~\ref{fig:amlagent_system_prompt} presents the system prompt used by AMLAgent.
The agent applies these principles to decide which accounts and transfers to investigate next as new records become available.
It provides fund-tracing guidance shared across cases and models, covering tracing directions, asset changes, and verification of candidate fund-flow connections.
It also guides the agent to select case-related transfers for its final submission.


\subsection{Task Prompt}

\begin{figure}[!t]
    \centering
    \includegraphics[width=\columnwidth]
        {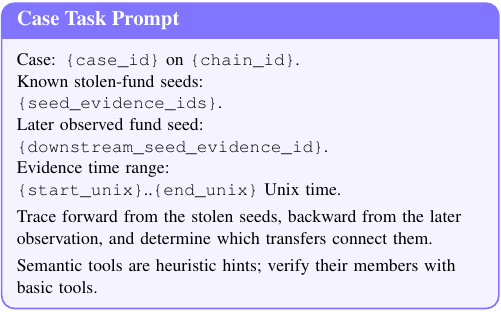}
      \vspace{-0.15in}
    \caption{Task prompt template of AMLAgent.}
    \label{fig:amlagent_task_prompt}
\end{figure}

Figure~\ref{fig:amlagent_task_prompt} presents the task prompt template used to initialize each investigation.
It specifies the case identifier, blockchain, investigation time window, and initial transfer identifiers, including downstream observations when available.
Fields in braces are replaced with case-specific values before each run.
The agent uses these initial clues as starting points for tool queries, progressively retrieving additional accounts and transfers during the investigation.


\section{Extended Experimental Evaluation (\S\ref{sec-evalua})}
\label{app:ex_experiment}

This section supplements the main evaluation with detailed
investigation costs across all 24 cases.
We also extend the four-model comparison to three more
cases to examine differences in tracing performance and
resource usage.

\subsection{Per-Case Investigation Costs}

Table~\ref{tab:investigation_cost} reports the investigation costs
of GPT-6 Astra across 24 cases.
Token usage ranges from 0.09M to 7.13M, with a mean of 1.73M,
indicating substantial variation across cases.
Bybit, Upbit, and PlusToken consume 7.13M, 4.36M, and 3.94M
tokens, respectively.
Together, these three cases account for approximately 37\%
of the total token usage, showing that a small subset of cases
contributes a substantial share of the overall resource consumption.

The three cost measures also reveal different patterns.
Fake Metadium and Harvest Finance both require 17 investigation
rounds, but use 213 and 45 tool calls and consume 1.62M and
1.98M tokens, respectively.
Thus, the same number of rounds can involve substantially
different query activity, and more tool calls do not necessarily
incur higher token usage.
Rounds capture decision steps, tool calls quantify tool
interactions, and token usage measures cumulative model input
and output.
Reporting these measures together provides a more complete
account of investigation costs than any single measure alone.

\subsection{Multi-Model Evaluation on Extra Cases}
\label{app:multi}
We further compare the four models on Cashio, Uranium Finance,
and SpankChain, with three runs per model and case.
Table~\ref{tab:small_case_model_comparison} reports the mean results.

\smallskip
\noindent\textbf{Recall loss primarily occurs during final selection.}
Of the 36 runs, 31 retrieve all annotated case-related transfers.
Across all runs, 97.63\% of missed transfer instances have already
appeared in tool responses but are not included in the final
submission.
Cashio further illustrates the types of omissions:
all 12 runs recover all 14 annotated transfers with nonzero
amounts from the first seed onward.
The recall differences mainly concern transfers with zero
amounts and activity preceding the first seed.

\smallskip
\noindent\textbf{Identical address recovery can mask differences
in transfer selection.}
Two Astra runs on SpankChain submit exactly the same 30
case-related transfers and involve identical address sets.
One run additionally submits 18 transfers outside the annotated
set, resulting in transfer-level F1 scores of 70.59\% and 89.55\%,
while their address-level metrics remain identical.
This shows that address metrics do not capture additional
transfers between addresses already included in the submission,
highlighting the need to combine transfer-level and address-level
evaluation when assessing fund-flow recovery.



\end{document}